\documentclass[12pt]{article}

\usepackage{times}
\usepackage{graphicx,epsfig,subfigure}    
\usepackage{mathptmx, textcomp} 
\usepackage{amsmath,amsthm,amssymb}              
\usepackage{rota ting}              

\title{ Dark state experiments with ultracold, deeply-bound triplet molecules}

\author
{Florian Lang,$^{1}$ Christoph Strauss,$^{1}$ Klaus Winkler,$^{1}$ Tetsu Takekoshi,$^1$
 \\ Rudolf Grimm,$^{1,2}$  Johannes Hecker Denschlag$^{1\ast}$ \\
\\
\normalsize{$^{1}$Institut f\"ur Experimentalphysik und Zentrum f\"ur Quantenphysik,} \\
\normalsize{ Universit\"at Innsbruck, A-6020 Innsbruck, Austria} \\
\normalsize{$^{2}$Institut f\"ur Quantenoptik und Quanteninformation} \\
\normalsize{ der \"Osterreichischen Akademie der Wissenschaften, A-6020 Innsbruck, Austria}\\
\\
\normalsize{$^\ast$ E-mail:  johannes.denschlag@uibk.ac.at} }

\date{}

\begin{document}

\baselineskip24pt

\maketitle

\begin{abstract}
We examine dark quantum superposition states of weakly
bound Rb$_2$ Feshbach molecules and tightly bound triplet Rb$_2$
molecules in the rovibrational ground state, created by subjecting a pure sample of Feshbach molecules in an optical lattice to a bichromatic Raman laser field. We analyze both experimentally and theoretically the creation and dynamics of these dark states. Coherent wavepacket oscillations of deeply bound molecules in lattice sites, as previously observed in Ref.~\cite{Lan08b}, are
suppressed due to laser-induced phase locking of molecular levels.
This can be understood as the appearance of a novel multilevel dark
state. In addition, the experimental methods developed help to
determine important properties of our coupled atom / laser system.

\end{abstract}

\section{Introduction}


Very recently, several groups have produced dense, ultracold
ensembles of molecules that are deeply bound \cite{Lan08b, Vit08, Dan08,
Ni08, Dei08} and in a ro-vibrational ground state
\cite{ Lan08b, Vit08, Ni08, Dei08}. This was achieved by binary
association of alkali atoms in ultracold ensembles via two
different pathways: (1) photoassociation \cite{Jon06, Hut06} and (2)
magneto-association at Feshbach resonances \cite{Hut06,Koh06}
combined with stimulated Raman adiabatic
passage (STIRAP)~\cite{Ber98}, a special coherent optical
transfer method. In contrast to photoassociation,
magneto-association only produces weakly-bound Feshbach molecules \cite{Hut06,Koh06}. STIRAP can then be used to transfer these weakly-bound molecules to the rovibrational ground state. This method is coherent, efficient, fast, reversible, and highly selective. STIRAP is based on a counter-intuitive light pulse sequence giving rise to a dynamically changing dark superposition state (Fig.~\ref{fig:potential}a)

\begin{equation}
|DS\rangle = \left({\Omega_2|f\rangle - \Omega_1|g\rangle}\right)/{\sqrt{\Omega_1^2+\Omega_2^2}}.
\label{equ:darkstate}
\end{equation}

In this paper, we deliberately replace the efficient but complex
STIRAP transfer of Ref.~\cite{Lan08b} with a simple square laser
pulse scheme. This reveals interesting
fundamental processes and dynamics in the coupled atom / laser system,
that would otherwise be hidden. In addition, this procedure
allows us to determine important
properties and parameters of our system and to check for consistency
with our theoretical model. We study the creation and
lifetime of dark superposition states that contain a sizeable
fraction of deeply bound molecules. These molecules are held in a 3D
optical lattice. Because the lattice potential is much shallower for the
deeply bound molecules than for the Feshbach molecules, and because the transfer is fast, the deeply bound molecules coherently populate several Bloch bands. In contrast to Ref.~\cite{Lan08b}, where similar circumstances lead to coherent
oscillations in the lattice, oscillations are suppressed in the
experiment described here due to phase locking of all
quantum levels involved. A novel dark state appears which is a
superposition of up to 8 quantum levels. We investigate the
limiting conditions under which oscillations set in.

\section{Experimental setup and initial preparation of molecules}

We carry out our dark state experiments with a 50\,$\mu$m-size pure ensemble of $3\times10^4$ weakly bound Rb$_2$ Feshbach molecules.
The molecules are trapped in the lowest Bloch band of a cubic 3D
optical lattice with no more than a single molecule per lattice
site~\cite{Tha06} and an effective lattice filling factor of about
0.3. The lattice depth for the Feshbach molecules is 60\,$E_r$,
where $E_r=\pi^2\hbar^2/2m a^2$ is the recoil energy, with $m$ the
mass of the molecules and $a = 415.22$\,nm the lattice period. Such
deep lattices suppress tunneling between different sites. A pure
ensemble of Feshbach molecules has been produced as follows. We
prepare a cold cloud of $6\times 10^5$ $^{87}$Rb atoms that are
either Bose condensed or \emph{nearly}\footnote{It turns out
that this increases the number of Feshbach molecules.} Bose condensed in a
Ioffe-type magnetic trap with trap frequencies $\omega_{x,y,z} = 2
\pi  \times (7, 19, 20)$ Hz. Within 100\,ms we adiabatically load the atoms
into the 3D optical lattice. After turning off the magnetic trap, we
flip the spins of our atoms from their initial state $| F = 1, m_F =
-1 \rangle$ to $|F = 1, m_F = +1 \rangle$ by suddenly reversing the
bias magnetic field of a few G. This spin state features a 210\,mG-wide Feshbach resonance at 1007.40\,G~\cite{Vol03}. By
adiabatically ramping over this resonance, we efficiently convert
atoms at multiply occupied lattice sites into Rb$_2$
Feshbach molecules. After conversion, inelastic collisions occur at
lattice sites that contain more particles than a single Feshbach
molecule, leading to vibrational relaxation of these molecules,
release of binding energy into kinetic energy and removal of all
particles from these sites. A subsequent combined microwave
and optical purification pulse removes all remaining chemically
unbound atoms, creating a pure sample of $3 \times 10^4$
Feshbach molecules. Afterwards, the magnetic field is set to
1005.8\,G, where the Feshbach molecules are in a quantum state
$|f\rangle$ which correlates with $|F=2, m_{F}=2, f_{1}=2, f_{2}=2,
v =36, l=0\rangle$ at 0\,G. Here, $F$ and $f_{1,2}$ are the total
angular momentum quantum numbers for the molecule and its atomic
constituents, respectively, and $m_{F}$ is the total magnetic
quantum number; $v$ is the vibrational quantum number for the
triplet ground state potential ($a ^3\Sigma_u^+$) and $l$ is the
quantum number for rotation.

\begin{figure}[t]
\includegraphics[width=0.85\textwidth]{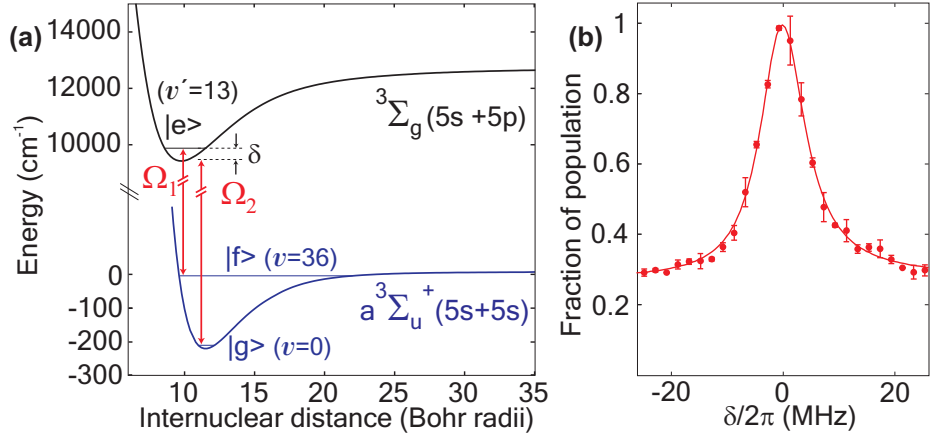}
\centering \caption{(a) $\Lambda$-type three-level scheme for dark states. The lasers 1 and 2 couple the molecular levels $|f\rangle$, $|g\rangle$ to the
excited level $|e\rangle$ with Rabi frequencies $\Omega_{1,2} $,
respectively.
(b) Typical dark resonance. Shown is the fraction of Feshbach molecules $|f\rangle$ remaining after exposing them to both lasers in a 3\,$\mu$s square pulse for varying two-photon detuning $\delta$. The data is identical to that shown in Ref.
\cite{Lan08b}.
}\label{fig:potential}
\end{figure}%

The bichromatic Raman laser field for the creation of the molecular
dark states is based on two lasers (1 and 2) which connect the
Feshbach molecule level $|f\rangle$, via an excited level
$|e\rangle$, to the absolute lowest level in the triplet potential
$|g\rangle$ (Fig.\,\ref{fig:potential}a). Laser 1 is a
Ti:Sapphire laser and laser 2 is a grating-stabilized diode laser.
Both lasers are Pound-Drever-Hall locked to a single cavity which
itself is locked to an atomic $^{87}$Rb-line. From the lock error
signals, we estimate frequency stabilities on a ms-timescale of
40\,kHz and 80\,kHz for lasers 1 and 2, respectively. Both laser
beams have a waist of 130\,$\mu$m at the location of the molecular
sample, propagate collinearly, and are polarized parallel to the
direction of the magnetic bias field. Thus, the lasers can only
induce $\pi$ transitions.

The ground state $|g\rangle$ has a binding energy of 7.0383(2)\,THz$\times h$
and can be described by the quantum numbers $|F=2, m_{F}=2, S=1,
I=3, v=0, l=0\rangle$ where $S$ and $I$ are the total electronic and
nuclear spins of the molecule, respectively.  At 1005.8\,G $|g\rangle$ is separated by hundreds of MHz from any other bound
level, so that there is no ambiguity as to which level is addressed. The
level $|e\rangle$ is located in the vibrational $v=13$ manifold of
the electronically excited $^3\Sigma_g$ (5s + 5p) potential and
has $1_\textrm{g}$ character. It has an excitation energy of
294.6264(2)\,THz$\times h$ with respect to $|f\rangle$, and a width
$\Gamma$ = $2\pi \times$ 8\,MHz. The Rabi frequencies $\Omega_{1,2}$
of the two lasers depend on their respective intensities
$I_{1,2}$,  i.\,e., $ \Omega_{1}=2\pi\times$0.4\,MHz$ \sqrt{I_1 /
(\mbox{W} \mbox{cm}^{-2})}$ and $ \Omega_2 =2\pi\times$30\,MHz
$\sqrt{I_2 / (\mbox{W} \mbox{cm}^{-2})}$, and are typically chosen
to be in the MHz regime.

\section{Dark state evolution within a square pulse}
\label{sec:darkstate}

\begin{figure}[t]
\includegraphics[width=0.65\textwidth]{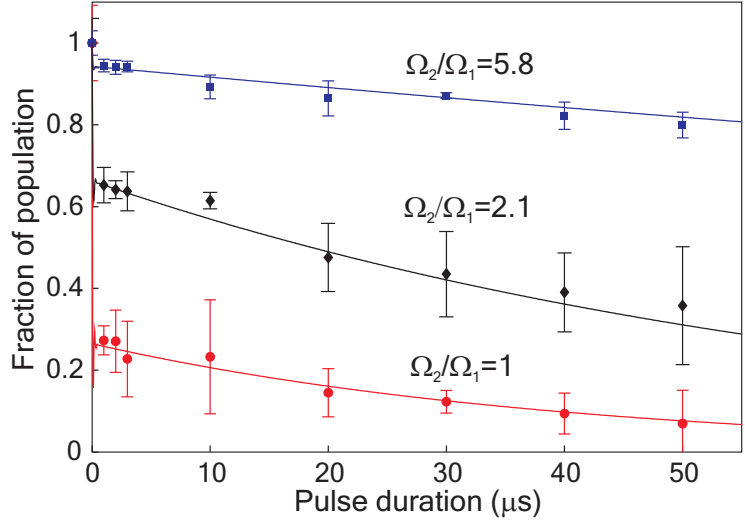}
\centering \caption{Dark state formation and lifetime. Shown is the
fraction of Feshbach molecules remaining after subjecting them to a
square pulse of Raman laser light of varying length for various Rabi frequency ratios $\Omega_{2}/\Omega_{1}$
($\Omega_{2}\approx2\pi\times7$\,MHz). After switching on the
lasers, a certain fraction of molecules is lost within 1\,$\mu$s and
a dark state has formed which has a much longer lifetime. The solid
lines represent model calculations (Sec.~\ref{sec:threelevelmodel}) which can be used to determine the Rabi
frequencies and short-term laser linewidths.}\label{fig:lifetime}
\end{figure}

Our square pulse projection experiments are carried out as follows. We expose the Feshbach molecules $| f \rangle$ in the lattice to square pulses of
Raman lasers 1 and 2 of variable pulse duration. Laser 2 is switched on
about 1\,$\mu$s before laser 1 to avoid excitation from $|f\rangle$ to $|e\rangle$ due to jitter in the laser pulse timing. The Raman lasers are resonant ($\delta = 0$) and the Rabi frequency $\Omega_{2}\approx2\pi\times7$\,MHz while $\Omega_{1}$ is
varied (Fig.~\ref{fig:lifetime}). After the pulse, we measure
the fraction of molecules remaining in state $|f\rangle$ by
dissociating them into pairs of atoms at the Feshbach resonance,
releasing them from the lattice and applying standard absorption
imaging. It is important to note that we
actually only count atoms in the lowest Bloch band of the lattice.
The release from the optical lattice is done as described
in \cite{Win06}, where after 13\,ms of ballistic expansion we
map out the Bloch bands in momentum space (see Appendix for
details).

Figure \ref{fig:lifetime} shows the remaining fraction of molecules
in state $|f\rangle$ versus pulse duration. Within 1\,$\mu$s we
observe a rapid loss of molecules that depends on the ratio
$\Omega_2 / \Omega_1$. The remaining molecules are stable on a much
longer timescale. This can be understood in terms of formation of a
dark state $|DS\rangle$. We can write

\begin{equation}
 |f\rangle =
\left( {\Omega_2|DS\rangle + \Omega_1|BS\rangle}
\right) / {\sqrt{\Omega_1^2+\Omega_2^2}}
\label{equ:darkbright}
\end{equation}

 where

\begin{equation}
|BS\rangle = \left({\Omega_1|f\rangle + \Omega_2|g\rangle}\right)/{\sqrt{\Omega_1^2+\Omega_2^2}}
\label{equ:brightstate}
\end{equation}

is a bright state which quickly decays via
resonant excitation to level $|e\rangle$. The dark state remains after the lasers are switched on and can be detected as a fraction
$\Omega_{2}^4/(\Omega_{1}^2+\Omega_{2}^2)^2$ of molecules projected back to $|f\rangle$ after switching off the lasers\footnote{This fact can be used to conveniently calibrate the Rabi frequency ratio $\Omega_{1}/ \Omega_{2}$. We found good agreement with other calibration methods for the Rabi frequencies.}.

Also, after the pulse a fraction $\Omega_{1}^2\Omega_{2}^2/(\Omega_{1}^2+\Omega_{2}^2)^2$ of the
initial molecules are in state $|g\rangle$ with a maximum of 25\%
for $\Omega_{1} = \Omega_{2}$. Thus, a sizeable fraction of the
molecules can be coherently transferred to the ground state. Remarkably, this
transfer takes place in less than 1 $\mu$s! Such short transfer
times cause Fourier broadening, resulting in considerably reduced laser stability requirements.
In addition, due to the formation of a dark state, there is still a well-defined phase relation between the $|f\rangle$ and $|g\rangle$ molecules.

As can be seen from Fig.~\ref{fig:lifetime}, the dark state slowly
decays. Its lifetime is shortest for $\Omega_{1} = \Omega_{2}$,
where we measure it to be $\approx$50\,$\mu$s. The decay of the dark
state is likely due to phase fluctuations of the Raman lasers. Phase
fluctuations lead to an admixture of a bright state component to the
otherwise dark state, which causes losses. In Sec.~\ref{sec:threelevelmodel} we will show that these fluctuations can
be expressed in terms of the short-term relative linewidth of the
lasers, $\gamma$, which we find to be about $2\pi \times 20$\,kHz. In
principle, the decay of the dark state could be due
to other effects, such as coupling to levels other than
$|f\rangle$, $|e\rangle$, and $|g\rangle$. However, we have verified
that this is not the case, because losses due to optical excitation
are completely negligible on the 100\,$\mu$s-timescale when we
expose a pure ensemble of $|f\rangle$ ($|g\rangle$) molecules to only
laser 2 (1).

We also searched for laser power dependent
shifts of the two-photon resonance. Using the Raman square
pulse measurements, we scanned the relative detuning of the lasers
for a fixed pulse duration and various laser powers. Within the accuracy
of our measurements of $2\pi\times200$\,kHz, we could not detect any
shifts of the resonance.

The behavior in Fig.~\ref{fig:lifetime} is described well by a
closed three-level model (a $\Lambda$ system) and its dynamics can
be simulated with a master equation which we describe in the
following.

\section{Three-level model and master equation}
\label{sec:threelevelmodel}
\newcommand{\ro}{\rho}
\newcommand{\si}{\sigma}

\begin{figure}[t]
\includegraphics[width=0.6\textwidth]{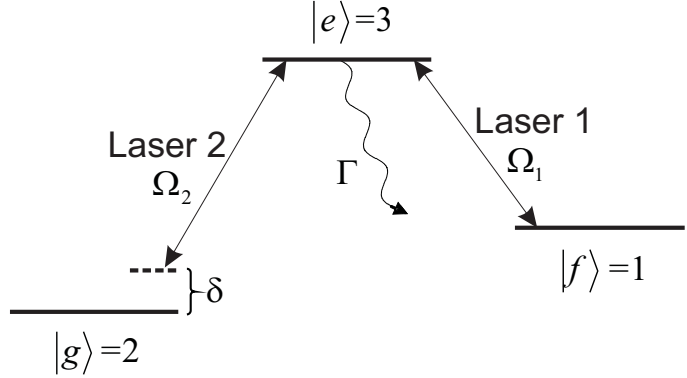}
\centering \caption{Level scheme for the master equation.}
\label{fig:st1}
\end{figure}

Neglecting lattice effects, we can describe the internal dynamics of
the molecules as they are subjected to the Raman laser fields with a
three-level model. We use a master equation~\cite{Wal94, Has88} which takes into account decoherence due to
phase fluctuations of the Raman lasers.
We consider the case where laser 1 is kept on resonance and laser 2
has a detuning~$\delta$ (Fig.~\ref{fig:st1}). Identifying the
levels  $|f\rangle$, $|g\rangle$, $|e\rangle$ with numbers 1,
2, 3, respectively, we can write the master equation as,

\begin{equation}
\begin{split}
\frac{d \ro}{dt} &= -i \delta \left[ \si_{}^{22}, \ro \right] - {i
\over 2}\sum_{k=1}^2 \Omega_k \left[ \si_{-}^{3k} + \si_{+}^{3k},
\ro \right] \\
&- {1 \over 2}\Gamma \left(
\si^{33} \cdot \si^{33} \cdot \ro + \ro \cdot \si^{33} \cdot \si^{33} \right)\\
&+   {1 \over 2} \gamma \left( 2 \si^{22} \cdot \ro \cdot
\si^{22} -\si^{22} \cdot \ro -\ro \cdot \si^{22} \right) ,
\end{split}
\end{equation}

where $\ro$ is the density matrix, $\Omega_{1,2}$ are the Rabi
frequencies, $\Gamma$ is the spontaneous decay rate of the excited
level $|e\rangle$, and $\gamma$ is the relative linewidth of the two
Raman lasers. The matrices $\si_{-}^{rs}$ and $\si_{+}^{rs}$ are
ladder operators and each is the transpose of the other. For example

\begin{equation}
\si_{-}^{32} = \left( \begin{array}{ccc}
0 & 0 & 0 \\
0 & 0 & 0 \\
0 & 1 & 0
\end{array}
\right) = \left(\si_{+}^{32}\right)^{\textrm{T}}.
\end{equation}

Setting the linewidth of the excited level $\Gamma$ = 8~MHz, the
detuning $\delta =  0$ and Rabi frequencies $\Omega_{2} = 2\pi\times
7$~MHz and $\Omega_1$ to give the ratios in Fig.~\ref{fig:lifetime}, we
fit all the data with a single fit
parameter $\gamma$. As a best fit, we obtain a relative linewidth of
the two Raman lasers $\gamma = 2 \pi \times 20$\,kHz, which is a
reasonable value for our laser system.

\section{Coherent oscillations and their suppression}
\label{sec:oscillations}

\begin{figure}[t!]
\includegraphics[width=0.5\textwidth]{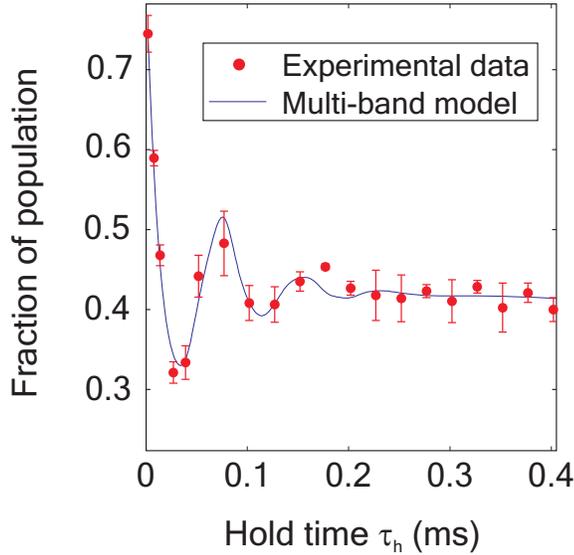}
\centering \caption{ We plot the transfer efficiency for the round-trip STIRAP process as a function of the hold time $\tau_h$ between
the two STIRAP pulses. With our procedure we only count molecules whose constituent atoms end up in the lowest Bloch band after transfer.
The oscillations in the transfer efficiency are due to breathing oscillations of localized spatial wavepackets of molecules in the lattice sites. The solid line is from a multi-band model calculation (Sec.~\ref{sec:model}). This plot is taken from Ref.~\cite{Lan08b}.} \label{fig:lifetime_gs}
\end{figure}

In reference~\cite{Lan08b} coherent oscillations of molecular
wavepackets of $|g\rangle$ molecules in the optical lattice were observed. We now investigate how these
observations fit together with the experimental results of the square pulse projection experiments presented here. For clarity, the oscillation data
from Ref.~\cite{Lan08b} are presented again in Fig.~\ref{fig:lifetime_gs} and briefly discussed.

\begin{figure}[tbh]
\includegraphics[width=0.7\textwidth]{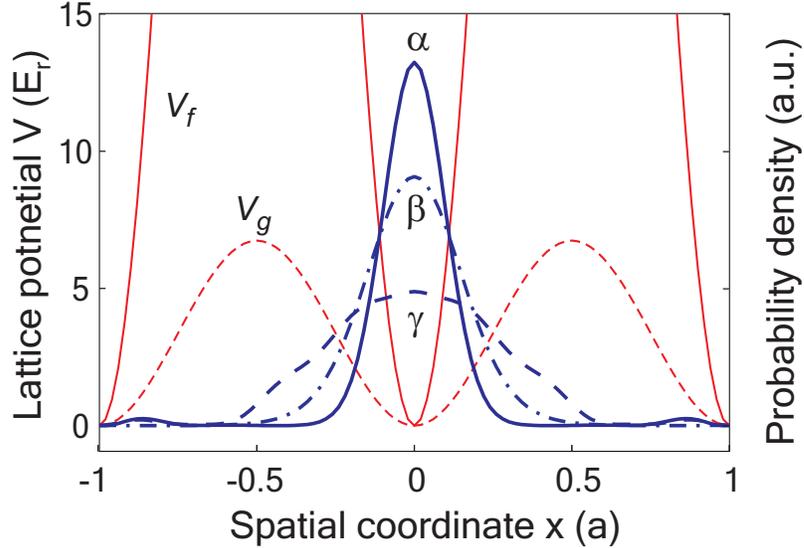}
\centering \caption{Wavepacket dynamics. Directly after the STIRAP transfer of a molecule from $|f\rangle$ to $|g\rangle$ the shape of its wavepacket (thick solid line $\alpha$) essentially corresponds to the vibrational ground state of the sinusoidal lattice potential for Feshbach molecules $V_f$ (thin solid line). In the much weaker potential felt by the ground state molecules $V_g$ (thin dashed line) the wavepacket starts to oscillate. After 1/8 of the oscillation period $\tau_{osc} = 2\pi/\omega_t$ its shape roughly corresponds to the vibrational ground state for $|g\rangle$ (thick dashed line $\beta$) and reaches its maximum extension after $\tau_{osc}$/4 (thick dash-dotted line $\gamma$).}

\label{fig:wavefunc_oscill}
\end{figure}

Using an STIRAP pulse sequence, Feshbach molecules are
efficiently transferred to level $|g\rangle$. The Raman lasers are extinguished and the molecules are held for a time $\tau_h$, after which they are transferred back to $|f\rangle$
with a reverse STIRAP pulse. The number of recovered Feshbach
molecules is counted. However, we only detect atoms that end up in
the lowest Bloch band after dissociation of the Feshbach molecules (see Appendix). The oscillation can be
understood as follows. We consider the localized spatial center-of-mass (c.o.m.) wavepacket of a Feshbach molecule at a particular lattice site in the lowest
Bloch band. The first STIRAP transfer projects this wavepacket onto
the much shallower\footnote{Due to a smaller dynamic polarizability, the lattice depth for the tightly bound $|g\rangle$ molecules is shallower than for the Feshbach molecules by a factor of $\approx$10.} lattice potential felt by the $|g\rangle$ molecules
(Fig.~\ref{fig:wavefunc_oscill}) without changing its shape. As
a consequence, $|g\rangle$ molecules are coherently spread over
various Bloch bands, and the wavepacket undergoes ``breathing''
oscillations with the lattice site trap frequency $\omega_t$. These
coherent oscillations (period $\approx$80\,$\mu$s ) are damped by
tunneling of $|g\rangle$ molecules in higher Bloch bands to
neighboring lattice sites. The reverse STIRAP transfer maps this
periodic oscillation back to the Feshbach molecule signal in Fig.~\ref{fig:lifetime_gs}. Higher Bloch
bands are populated here as well, but are at most partially counted
in our scheme (see Appendix), which leads to an apparent decrease in
transfer efficiency.

The question arises why similar oscillations are not observed in our
square pulse projection measurements shown in Fig.~\ref{fig:lifetime},
especially for the case $\Omega_1 = \Omega_2$ where 50\% of the
population is in state $|g\rangle$. One might assume that the
spatial wavepackets of the $|g\rangle$ molecules undergo similar
breathing oscillations. These oscillations would then periodically break up the dark superposition state and lead to corresponding losses. They would also
periodically produce population in higher Bloch bands of the
Feshbach molecule lattice. As we will see, the oscillations are suppressed because the Raman lasers phase lock the involved quantum levels which stops, in a sense, the free evolution of the wavepackets. We can understand this
behavior in detail with the help of a multi-band model, which we
describe in the following.

\section{Multi-band model}
\label{sec:model}

\begin{figure}[t]
\includegraphics[width=0.65\textwidth]{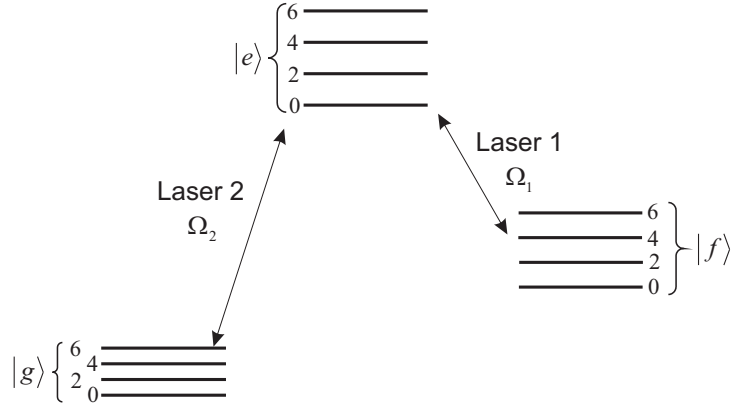}
\centering \caption{Multi-band model. The three molecular levels
$|f\rangle$, $|e\rangle$ and $|g\rangle$ have a Bloch band substructure
due to the optical lattice. We restrict the model to
the 4 lowest Bloch bands with even symmetry (band index $n = 0, 2,
4, 6$).
} \label{fig:st2}
\end{figure}

In an optical lattice the molecular levels $|f\rangle$, $|g\rangle$
and $|e\rangle$ from the previous model have a
substructure given by the lattice Bloch bands. Because the lattice
depths for the levels $|f\rangle$, $|g\rangle$ and $|e\rangle$ are in
general different, the respective band structures will also vary.
 This combination of external
(c.o.m. motion in the lattice) and internal degrees of
freedom gives rise to a number of new quantum levels which
 are coupled by the laser fields (Fig.~\ref{fig:st2}). We assume each Feshbach molecule to be
initially localized in a singly-occupied lattice site. The
corresponding localized molecular wavepacket can be described by
Wannier functions~\cite{Koh73} which form a complete set of
orthonormal functions. In the following we will denote the Wannier
function for level $|\alpha\rangle$ and band $n$ as $|\Psi_{\alpha n}\rangle$.
We note that for deep lattices, these Wannier functions
closely resemble harmonic oscillator wavefunctions.


The Raman lasers couple different $|\Psi_{\alpha n}\rangle$ according to
the respective wavefunction overlaps (Fig.~\ref{fig:st2}). Since
the initial wavepackets of the Feshbach molecules are symmetric,
only even bands will be populated. We restrict our calculations to
the four lowest Bloch bands with even symmetry,\footnote{The effect
of including higher bands into the model was found to be
negligible.} corresponding to the band indices $n = 0, 2, 4, 6$. The
dynamics in each of the three lattice directions is then described by a
12-level model, which can in principle be solved in terms of a
master equation (Sec.~\ref{sec:threelevelmodel}). However, we have
used a Schr\"odinger equation-based model since the numerical code
is less involved. In this approach, laser phase fluctuations are
neglected, and we introduce a lattice site tunnel rate for each
band. These tunnel rates are chosen to match the expected tunnel rates for the different bands and are slightly adjusted for a better
fit of the data in Fig. \ref{fig:lifetime_gs}. We note that the
results of the model calculations are essentially independent of the
excited state lattice depth, which is not well known.

The Hamiltonian $H$ of our time dependent Schr\"{o}dinger equation
\begin{equation}
i\hbar\frac{\partial}{\partial t}|\Phi\rangle = H|\Phi\rangle
\label{equ:schroedinger}
\end{equation}
has the form of a $12\times12$ matrix,
\begin{equation}
H = \hbar \left( \begin{array}{ccccc}
E_{f0} - \frac{i}{2} J_{f0} & 0 & \frac{1}{2} \Omega_1(t) \cdot M_{f0,e0} & 0 & \dots \\
0 & E_{g0} + \delta - \frac{i}{2} J_{g0}  & \frac{1}{2} \Omega_2(t) \cdot M_{g0,e0} & 0 & \dots \\
\frac{1}{2} \Omega_1(t) \cdot M_{e0,f0} & \frac{1}{2} \Omega_2(t) \cdot M_{e0,g0} & E_{e0} - \frac{i}{2} \Gamma - \frac{i}{2} J_{e0} & \frac{1}{2} \Omega_1(t) \cdot M_{e0,f2} & \dots \\
0 & 0 & \frac{1}{2} \Omega_1(t) \cdot M_{f2,e0} &  E_{f2} - \frac{i}{2} J_{f2} & \dots \\
\vdots & \vdots & \vdots & \vdots & \\
\end{array}
\right).
\end{equation}

Here $E_{\alpha n}$ and $J_{\alpha n}$ are the energy 
and tunnel matrix element respectively for the Wannier function
$|\Psi_{\alpha n}\rangle$ in band $n$ of level $|\alpha \rangle$. $M_{\alpha n, \beta k} =
\langle\Psi_{\alpha n}|\Psi_{\beta k}\rangle$ is the overlap integral of the
respective Wannier functions.

\begin{figure}[thb]
\centering
\subfigure{\includegraphics[width=0.6\textwidth]{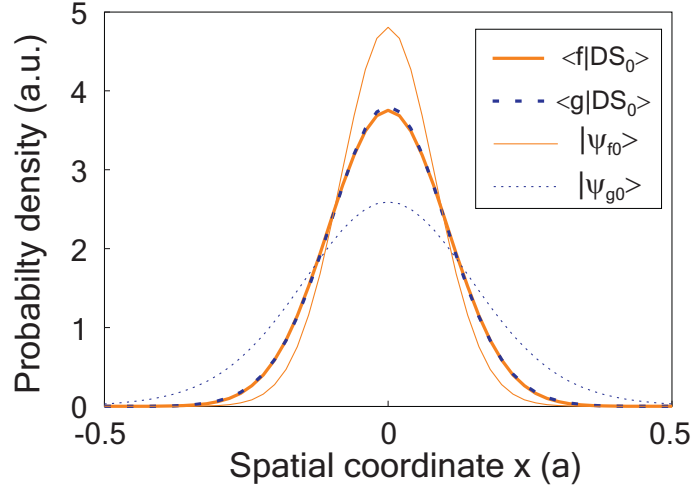}}
\caption{
Spatial wavepackets of the dark state $|DS_{0}\rangle$
and the Wannier functions $|\Psi_{f0}\rangle$, $|\Psi_{g0}\rangle$.
The dark state $|DS_{0}\rangle$ has two components, one having
$|f\rangle$ character ($\langle f|DS_{n}\rangle$) and the other one
having $|g\rangle$ character ($\langle g|DS_{n}\rangle$). The
wavepackets of $\langle f|DS_{n}\rangle$ and $\langle
g|DS_{n}\rangle$ essentially have the same shape. They are
mainly composed of the lattice ground states $|\Psi_{f0}\rangle$ (thin solid
line) and $|\Psi_{g0}\rangle$ (thin dotted line). All depicted
states are normalized. The parameters used are
$\Omega_1=\Omega_2=2\pi\times7$~MHz, $V_f$ = 60\,$E_r$ and $V_g$ =
6\,$E_r$, as in our experiments.} \label{fig:darkstates}
\end{figure}

Diagonalizing this Hamiltonian, we find twelve ``eigenstates'' of
the coupled system which in general have complex eigenvalues. In the
following, we study the case of strong coupling ($\Omega_{1,2}\gg \omega_t),$\footnote{For our experiments where $\Omega_{1,2}\gtrsim 2\pi\times 1$\,MHz and $\omega_{t}\sim 2\pi\times 10$\,kHz this condition is satisfied.} which is the regime for phase locking.
In this regime, four of these eigenstates have negligible
contribution from the exited level $|e\rangle$ and thus a long
lifetime. These 4 quasi-dark states essentially correspond to the 4
lattice bands in our model and will be denoted as $|DS_{n}\rangle$
with n = 0, 2, 4,~6. We now study the spatial waveforms of these
dark states (Fig.~\ref{fig:darkstates}) and compare the components
with $|f\rangle$ and $|g\rangle$ character. Neglecting a small $|e\rangle$ component the dark superposition state, $|DS_{n}\rangle$ can be written as
\begin{equation}
|DS_{n}\rangle = |g\rangle \langle g|DS_{n}\rangle + |f\rangle
\langle f|DS_{n}\rangle.
\end{equation}
As an example ($\Omega_1 = \Omega_2$) Fig.~\ref{fig:darkstates} shows that the wavepackets $\langle g|DS_{n}\rangle$ and $\langle f|DS_{n}\rangle$ have
the same shape. This is not surprising since this ensures that the ratio of the $|f\rangle$ and $|g\rangle$ amplitudes equals $\Omega_2/\Omega_1$ everywhere, as in Eq.~\ref{equ:darkstate}.


%
%


Let us now discuss the formation and evolution of the dark state
that we have observed in the square pulse experiments of Sec.~\ref{sec:darkstate}. A dark state $|DS\rangle$ is
formed in less than 1\,$\mu s$ by subjecting Feshbach molecules to a
square Raman laser pulse. As in the STIRAP transfer (discussed in
Sec.~\ref{sec:oscillations}) the initial projection onto $|DS\rangle$ will not change the shape of the Feshbach molecule wavepacket, given by the Wannier function $|\Psi_{f0}\rangle$. The dark state  can be expressed as a coherent superposition of the four dark eigenstates $|DS_n\rangle$ of the 12-level Hamiltonian
\begin{equation}
|DS\rangle = \sum_{n=0,2,4,6}{c_n|DS_n\rangle}.
\label{equ:darkstate_n}
\end{equation}
The subsequent coherent evolution of these dark states will again in principle lead to breathing oscillations. The amplitude of these oscillations depends on the extent to which higher bands (i.\,e., $|DS\rangle_n$, $n>0$) are excited. The excitation increases with increasing deviation of $|DS\rangle$ from the initial state $|f\rangle$, i.\,e., with rising $\Omega_1/\Omega_2$.

This can also be understood from another point of view. The effective
lattice potential felt by the molecules in such a superposition
state is the weighted average of the potentials for the two contributing
states $|f\rangle$ and $|g\rangle$.
For the case $\Omega_1=\Omega_2$ this effective potential is about half as deep as that for the Feshbach molecules. Compared to the case of pure ground state molecules (Fig.~\ref{fig:lifetime_gs}) where the lattice potential is reduced by a factor of 10, the oscillations of the wavepacket are strongly suppressed and cannot be observed with our current experimental precision. For $\Omega_1\gg\Omega_2$, the dark state $|DS\rangle$ has a dominant contribution from state $|g\rangle$, and the effective
lattice potential essentially corresponds to the one for ground
state molecules. In this case oscillations appear despite
the strong coupling, a fact which we also have experimentally
verified\footnote{For these experiments we have ramped into the dark
state and back in a fashion similar to STIRAP transfer pulses to
avoid strong losses caused by direct projection into $|DS\rangle \approx |g\rangle\langle g|DS\rangle$.}.

\paragraph*{Conclusion}

We have analyzed coherent wavepacket dynamics and their
suppression in a 3D optical lattice. We observed optically induced
phase locking of a number of quantum levels, which can also be
viewed as the appearance of a novel multi-level dark state. The
experiments were carried out with tightly bound molecules as a
component of a dark quantum superposition state. Thus, the
experiments demonstrate
control of molecular motion in an optical lattice for the first time. In addition,
different models have been introduced and discussed in detail, with
which the lattice dynamics can be understood and quantitatively
described.

%
%

\paragraph*{Acknowledgements}

The authors thank Helmut Ritsch for very helpful discussions
 and support in model calculations.
  We also thank Gregor Thalhammer for early assistance in the lab,
  and Florian Schreck for loaning us a Verdi V18 pump laser.
  This work was supported by the Austrian Science Fund (FWF) within SFB 15 (project part 17).

\section*{Appendix: Theoretical band population analysis}

\begin{figure}[t]
\includegraphics[width=0.4\textwidth]{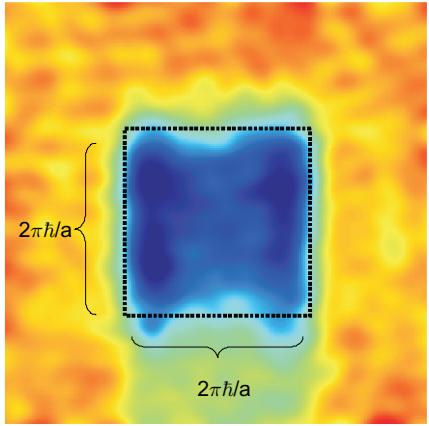}
\centering \caption{Shown is a typical absorption image which
displays the atomic quasi-momentum distribution in the optical
lattice after exposure to the Raman laser beams and
subsequent adiabatic molecule dissociation. Atoms inside the
square region come from the lowest Bloch band. $2\pi\hbar/a$ is the
modulus of the reciprocal lattice vector.
} \label{fig:banddist}
\end{figure}

As stated before, our signals only include molecules for which
the constituent atoms end up in the lowest Bloch band of the
lattice. A controlled lattice rampdown in a few milliseconds
maps the bands and quasi-momentum distribution of the atoms into
momentum space \cite{Den02,Win06}. We image these distributions
after 13~ms of time-of-flight via absorption imaging. Fig.
\ref{fig:banddist} shows a typical distribution.
The dotted square region corresponds to the lowest Bloch band and is
dominantly populated.

\begin{figure}[t]
\includegraphics[width=0.85\textwidth]{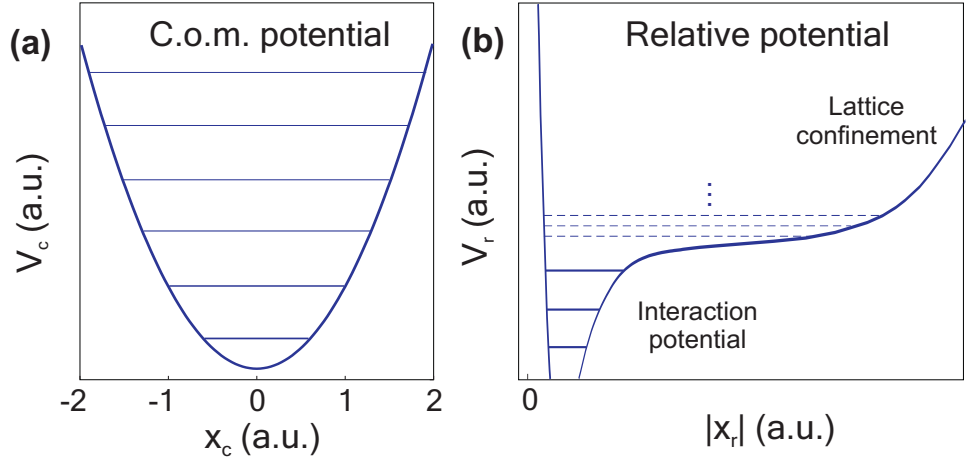}
\centering \caption{Potentials for the center-of-mass and relative
coordinate of two atoms trapped at a site of the optical lattice.
(a) The bound states (solid lines) of the harmonic center-of-mass
potential $V_c$ correspond to the molecular Bloch bands. (b) At
short interatomic distances the relative potential $V_r$ is
dominated by the interaction potential, which allows the formation
of bound molecular states (solid lines). Unbound atoms are trapped
by the lattice potential at larger separation (dashed lines). With the
help of a Feshbach resonance the lowest trap state can be converted
into a high molecular state. Note that in this schematic view
both energy and distance for the two contributions to the relative potential are not to scale.}
\label{fig:lattice_pot}
\end{figure}

An important question is how the Bloch bands for the Feshbach
molecules map onto the Bloch bands for the atoms. In other words, if
we measure the atomic population of the Bloch bands -- do we know
what the band population for the molecules was? As the lattice is very deep for the Feshbach molecules and atoms, we can approximate the potential at an individual lattice site as harmonic with trap frequency $\omega_t$. In one dimension, the eigenfunctions of the harmonic oscillator are \begin{equation} |\Phi_n\rangle = {1 \over \sqrt{2^n n! \sqrt{\pi} x_0}} \ \exp\left( - {1 \over 2} \left({x \over x_0}\right)^2 \right) \ \mbox{H}_n\left({x \over x_0}\right),
\end{equation}

where $x_0 = \sqrt{\hbar / \omega_t m}$ is the oscillator length and H$_n$ is the n$^{\textrm{th}}$ Hermite polynomial.
We assume that we have two atoms in a lattice site with coordinates
$x_{1,2}$. The relative and c.o.m. coordinates of the
atom pair are
\begin{eqnarray}
x_r &=& 1/\sqrt{2} (x_1 - x_2) \\
x_c &=& 1/\sqrt{2} (x_1 + x_2)
\end{eqnarray}
The c.o.m.$\omega_t$. In one dimension, the eigenfunctions of the harmonic oscillator are
We assume that we have two atoms in a lattice site with coordinates
$x_{1,2}$. The relative and c.o.m. coordinates of the
atom pair are
\begin{eqnarray}
x_r &=& 1/\sqrt{2} (x_1 - x_2) \\
x_c &=& 1/\sqrt{2} (x_1 + x_2)
\end{eqnarray}
The c.o.m. potential $V_c$ for the pair will be harmonic
with trap frequency $\omega_t$
and the potential $V_r$ for the relative coordinate will be a sum of
the harmonic potential and the interaction potential (Fig.~\ref{fig:lattice_pot}).

When we form or dissociate a molecule by adiabatically ramping
across a Feshbach resonance, only the quantum level in the $V_r$
potential will change -- from a molecular bound state to an unbound atomic
pair state in the lowest Bloch band. The wavefunction in the c.o.m. coordinate remains unchanged. We can now calculate how band
populations of Feshbach molecules converted to atomic band populations
by using the coordinate transformations. As an example: A Feshbach
molecule in the lowest Bloch band (i.\,e., center-of-mass coordinate)
will produce an atom pair with the following wavefunction:
$|\Psi\rangle \propto \exp(-1/2 \ x_c^2) \exp(-1/2 \ x_r^2) =
\exp(-1/2 \ x_1^2) \exp(-1/2 \ x_2^2)$. This means that both atoms
will also end up in the lowest Bloch band of the lattice. This
analysis can be extended to any band. Table~\ref{tab:bandconver}
gives the conversion amplitudes from molecular to atomic bands for
the four lowest symmetric molecular bands. Correlations between the two constituent atoms of a molecule are not discussed here.

We finally note that when we apply absorption imaging, the optical
density of the atomic sample is integrated in the direction of
observation. Thus in this direction no band population analysis is
possible. We accounted for this in our multi-band model described in Sec.~\ref{sec:model}.

\cleardoublepage

\vspace{0.7cm}
\begin{table}[htb]
\centering
\begin{tabular}{|cc|ccccccccc|}
\hline
&&\multicolumn{8}{c}{atomic Bloch band}&\\
&&0&2&4&6&8&10&12&14&\dots \\
\hline
&0&1&0&0&0&0&0&0&0& \dots  \\
&2&$\frac{1}{2}$&$\sqrt{\frac{1}{2}}$&$\frac{1}{2}$&0&0&0&0&0& \dots \\
&4&$\frac{1}{4}$&$\sqrt{\frac{3}{16}}$&$\sqrt{\frac{1}{2}}$&$\sqrt{\frac{3}{16}}$&$\frac{1}{4}$&0&0&0& \dots \\
&6&$\frac{1}{8}$&$\sqrt{\frac{3}{64}}$&$\sqrt{\frac{3}{16}}$&$\sqrt{\frac{1}{2}}$&$\sqrt{\frac{3}{16}}$&$\sqrt{\frac{3}{64}}$&$\frac{1}{8}$&0& \dots \\
\hspace{3mm}\begin{rotate}{90}~~~molecular band\end{rotate}&\vdots&\vdots&\vdots&\vdots&\vdots&\vdots&\vdots&\vdots&\vdots&  \\
\hline
\end{tabular}
\caption{Band conversion amplitudes in the harmonic oscillator
approximation. Each line gives the amplitudes for a constituent atom
of a molecule in a certain band to populate various atomic
bands after dissociation.}
\label{tab:bandconver}
\end{table}

\cleardoublepage

\end{document}